\documentstyle[preprint]{jpsj}
\title{ Application of 
        the Density Matrix Renormalization Group Method
        to a Non-Equilibrium Problem
}
\author{Yasuhiro {\sc Hieida}
\footnote{e-mail: hieida@godzilla.phys.sci.osaka-u.ac.jp}
}
\inst{ Department of Physics,
       Graduate School of Science, Osaka University,\\
       Machikaneyama-cho 1-1, Toyonaka, Osaka 560, Japan\\
}
\abst{
We apply the density matrix renormalization group (DMRG) method to a
 non-equilibrium problem: the asymmetric exclusion process in one
 dimension. We study the stationary state of the process to calculate
 the particle density profile (one-point function).  We show that,
 even with a small number of retained bases, the DMRG calculation is
 in excellent agreement with the exact solution obtained by the
 matrix-product-ansatz approach.
}
\kword{
        asymmetric exclusion process,
        ASEP,
        density matrix renormalization group,
        DMRG,
        non-equilibrium phenomena,
        quadratic algebra
}
\begin{document}
\sloppy
\maketitle


\section{Introduction}

Many phenomena in the vast realm of non-equilibrium physics can be
viewed as the asymmetric (simple) exclusion process (ASEP for
short)~\cite{SZ,YKT}. Models of the ASEP in one spatial dimension have
been extensively studied since a remarkable paper~\cite{Derrida}
appeared, which proposed the so-called matrix product ansatz
(MPA). For a class of models, the exact stationary probability
distributions have been obtained by the MPA. However, there are many
models whose stationary probability distributions cannot be obtained
by using the MPA. Alternative approaches are required to study general
models of the ASEP.

There are two types of models for the ASEP: discrete-time models and
the continuous-time models. The discrete-time models which we treat in
this paper are more general than the continuous-time ones because the
latter are realized by taking suitable limits of the parameters in the
former.

The density-matrix renormalization group (DMRG) method invented by
 S. R. White is an approximate numerical scheme for diagonalization of
 Hamiltonians, which has been applied to various one-dimensional
 quantum systems, with results demonstrating its surprising
 efficiency~\cite{orig-DMRG}. As pointed out by T. Nishino, the DMRG
 is also applicable to two-dimensional lattice statistical models for
 diagonalization of the transfer matrices~\cite{Nishino}. Since the
 time-evolution operator in the discrete-time ASEP can be regarded as
 a transfer matrix, we can naturally expect applicability of the DMRG
 also to the ASEP. In the actual implementation, however, we should
 consider the following two points.  First, we should treat a
 non-symmetric transfer matrix. Second, the role of the wave function
 $\psi$ is different; in the ASEP not $|\psi|^2$ but $\psi$ itself is
 the probability distribution.

In this paper, by properly incorporating the above two points, we
apply the DMRG to a model of the ASEP to show that the DMRG is an
efficient numerical approach also to non-equilibrium problems.

\section{Model}

Let us explain the model which we treat. It is the model of stochastic
process of hopping particles on a chain of size $N$.  For technical
simplicity, we will assume $N$ to be $4 \times ({\rm integer}) + 2$.
The time is taken to be discrete.  Each site can have two states:
either empty or occupied.  Particles hop to the right-nearest site
with probability $p$ and to the left-nearest site with probability
$q$.  The chain are coupled to particle reservoirs at the both ends:
if the leftmost site is empty (resp. occupied), a particle is added
(resp. removed) with probability $\alpha$ (resp. $\gamma$); if the
rightmost site is empty (resp. occupied), a particle is added
(resp. removed) with probability $\delta$ (resp. $\beta$).

Time evolution of the model consists of the following two steps. \\

\noindent
[A-1]\\
\noindent
For all
 odd-even pairs of sites $\{(2j-1,2j)\}$ ($j=1,2,3,...,N/2-1$)
their states are updated simultaneously. States of the leftmost and
the rightmost sites are updated in this step.\\

\noindent
[A-2]\\
\noindent
For all
 even-odd pairs of sites $\{(2j,2j+1)\}$ ($j=0,1,2,...,N/2-1$)
 their states are updated simultaneously.\\

We introduce a transfer matrix $T$ which describes the discrete-time
evolution of the probability distribution. Corresponding to the two
steps of the time evolution of the model, $T$ consists of two transfer
matrices $T_1$ and $T_2$:
\begin{equation}
T=T_2 T_1\, ,
\end{equation}
where $T_1$ and $T_2$ correspond to the first step [A-1] and second
step [A-2] of the time evolution, respectively.  Explicitly,
\begin{equation}
    \begin{array}{rcl}
 T_1 & = &
{\cal L} \otimes {\cal T}_{1,2}\otimes {\cal T}_{3,4}
  \cdots  {\cal T}_{N-3,N-2} \otimes {\cal R}\\
 T_2 & = &
{\cal T}_{0,1}\otimes {\cal T}_{2,3}
 \cdots  {\cal T}_{N-2,N-1}\, ,
    \end{array}
\end{equation}
where ${\cal L}$ and ${\cal R}$ are the left boundary transfer matrix
 and the right boundary transfer matrix, respectively and ${\cal
 T}_{i,i+1}$ is a local transfer matrix acting on the pair of the
 sites $i$ and $i+1$.

In order to represent explicit matrix elements, we introduce a state
variable $i_n$ at each site $n$ as:
\begin{equation}
        i_n=0 \quad ({\rm empty}); \quad
        i_n=1 \quad ({\rm occupied}).
\label{eq:state-variables}
\end{equation}
Then, the local transfer matrix ${\cal T}$ is given by 
\begin{equation}
{\cal T}_{n,n+1}=
\bordermatrix{
      & (0,0) & (0,1) & (1,0) & (1,1) \cr
(0,0) &     1 &     0 &     0 &     0 \cr
(0,1) &     0 &   1-q &     p &     0 \cr
(1,0) &     0 &     q &   1-p &     0 \cr
(1,1) &     0 &     0 &     0 &     1 \cr
}
\label{eq:local-TM}
\end{equation}
where the element ${({\cal T}_{n,n+1})}_{(f_n,f_{n+1}),(i_n,i_{n+1})}$
represents the probability for the transition from the state $
(i_n,i_{n+1}) $ to the state $ (f_n,f_{n+1}) $.  The boundary transfer
matrices are
\begin{equation}
{\cal L}=
\bordermatrix{
  & 0        &1         \cr
0 & 1-\alpha & \gamma   \cr
1 & \alpha   & 1-\gamma \cr
}
\, , 
{\cal R}=
\bordermatrix{
  & 0        &1         \cr
0 & 1-\delta & \beta    \cr
1 & \delta   & 1-\beta  \cr
}
\end{equation}

The stationary state of the model has three phases: the high density
 phase, the low density phase and the maximal current phase. They are
 separated by the boundary-induced phase transitions~\cite{HP,SZ,BI}.

\section{Method}
In this article, we concentrate on the density profile (one-point
function) $\langle i_n \rangle$ ($0\le n \le N-1$) in the stationary
state, which can be calculated in two steps as follows.\\

\noindent
[B-1]\\
\noindent
Obtain the right largest-eigenvalue eigenvector $|P \rangle$ of $T$
which satisfy
\begin{equation}
T|P \rangle = \lambda |P \rangle \, ,
\label{eq:def-P}
\end{equation}
 with $\lambda(=1)$ being the largest eigenvalue of $T$. This $|P
\rangle$ is the stationary state whose wavefunction $\langle
i_0,i_1,i_2,\cdots,i_n,\cdots,i_{N-2},i_{N-1}|P \rangle$ gives the
stationary probability distribution.

\noindent
[B-2]\\
\noindent
 Calculate $\langle i_n \rangle$:
\begin{equation}
\langle i_n \rangle
=\frac
    {
    \sum_{\{i\}} i_n
        \langle i_0,i_1,i_2,
        \cdots,i_n,\cdots,i_{N-2},i_{N-1}|P \rangle
    }{
    \sum_{\{i\}}\quad
        \langle i_0,i_1,i_2,
        \cdots,i_n,\cdots,i_{N-2},i_{N-1}|P \rangle
} \, ,
\label{eq:def-tau}
\end{equation}
where $\sum_{\{i\}}$ denotes summation over all states
$\{i_0,i_1,\cdots,i_{N-2},i_{N-1}\}$ of $N$ sites.

We should note that to calculate the quantity $\langle i_n \rangle$,
 we do not use the left largest-eigenvalue eigenvector of $T$.  This
 contrasts with the case of the classical statistical system, where
 the one-point function is expressed as $\langle Q|\hat{i_n}|P\rangle
 / \langle Q|P\rangle$ with $\langle Q|$ being the left
 largest-eigenvalue eigenvector of $T$ and $\hat{i_n}$ the local
 density operator.

Let us explain how we can incorporate the DMRG in these processes
[B-1] and [B-2].  The method we employ is essentially the
finite-system-algorithm version~\cite{orig-DMRG} of the ``classical''
DMRG~\cite{Nishino}. However, since our transfer matrix $T$ is {\it
not} symmetric, we need a non-trivial modification in the actual
application of the algorithm: Unlike the ordinary DMRG, we should deal
with ``asymmetric density matrix''~\cite{CTMRG-Lett,CTMRG}.

To clarify our explanation, we treat a chain of size $N=14$. Suppose
that, at some iteration stage in the DMRG, the renormalized transfer
matrix $T$ is expressed as in Fig.~\ref{fig-14N}. For such form of
$T$, we adopt the notation ${T_4^l}{}_\times {}^\times {}_\times
{T_8^r}$, where $T_4^l$ is the renormalized left block transfer matrix
of 4 sites, ``$ \times $'' represents the local transfer matrix and
$T_8^r$ is the renormalized right block transfer matrix of 8
sites. Just as in the original finite-size algorithm of the DMRG, we
consider the set $\{{T_k^l}{}_\times {}^\times {}_\times
{T_{N-k-2}^r}\}_{k=2,4,\ldots,N-4}$ to refine iteratively the
largest-eigenvalue eigenstate.

Hereafter we use Greek indices to represent the block state.  By the
power method (i.e., multiplication of $T$ sufficiently many times), we
get the right eigenfunction
 $\psi(\mu_3; i_3,i_4,i_5,i_6; \nu_7) $
 and
the left eigenfunction
 $\phi(\mu_3; i_3,i_4,i_5,i_6; \nu_7) $
belonging to the largest eigenvalue of
 $T={T_4^l}{}_\times {}^\times {}_\times {T_8^r}$.  We then form
 the ``density matrix''~\cite{CTMRG-Lett,CTMRG,BXG,WX,Shibata}
 for the left 5 sites
 $\rho^l_5=\{\rho^l_5( \mu_3,i_3,i_4;
{{\mu}^\prime_3},{i^\prime_3},{i^\prime_4})\}$
 as
\begin{equation}
\rho^l_5(\mu_3,i_3,i_4;
{{\mu}^\prime_3},{i^\prime_3},{i^\prime_4})
\equiv
\sum_{i_5,i_6,\nu_7}
\psi({\mu}_3;i_3,i_4,i_5,i_6;{\nu}_7)
\phi({\mu}^\prime_3;
{i^\prime_3},{i^\prime_4},
i_5,i_6; {\nu}_7)
\, .
\end{equation}
Since $\phi \neq \psi$, in general, for asymmetric $T$, the matrix
$\rho^l_5$ is also asymmetric:
 $\rho^l_5(\mu_3,i_3,i_4;
{{\mu}^\prime_3},{i^\prime_3},{i^\prime_4})
 \neq
\rho^l_5({{\mu}^\prime_3},{i^\prime_3},{i^\prime_4};
 \mu_3,i_3,i_4)$. 
As in the ordinary DMRG,
 we must perform eigenvalue decomposition of $\rho^l_5$
 to choose a truncated basis set for a new
 5-site
 block. In
our case, we should obtain both of the right and left eigenvectors for
this purpose. Using the obtained transformation matrices for the new
block states, we can renormalize the left block transfer matrix of 6
sites ${T_4^l}{}_\times {}^\times {} $ to get $T_6^l$. All the
necessary block-making procedure can be done in a similar fashion.

By $\{L^j_{({\mu}_{j-2},i_{j-2},i_{j-1}),\mu_j}\}$
 we denote the components of
the $\mu_{j}$-th right eigenvector of $\rho^l_j$
 (density matrix for
left $j$ sites) with eigenvalue $w^l_{\mu_j}$:
\begin{equation}
\sum_{{\mu}_{j-2},{i_{j-2}},{i_{j-1}}}
\rho^l_j(
  {\mu}^\prime_{j-2},{i^\prime_{j-2}},{i^\prime_{j-1}};
  {\mu}_{j-2},{i_{j-2}},{i_{j-1}})
L^j_{({\mu}_{j-2},i_{j-2},i_{j-1}),\mu_j}
= 
w^l_{\mu_j}
L^j_{
  (
    {\mu}^\prime_{j-2},
    {i^\prime_{j-2}},
    {i^\prime_{j-1}}
  ),\mu_j} \quad ,
\end{equation}
 where we set $\mu_1\equiv i_0 $.
For the density matrix $\rho^r_j$ for right $j$ sites, we denote its
right eigenvector with eigenvalue
 $w^r_{\nu_j}$
 by $R^{j}$,
 whose
components
 $\{R^j_{(i_{N-j},i_{N-j+1},{\nu}_{j-2}),\nu_j}\}$
 satisfy
\begin{eqnarray}
&\sum_{i_{N-j},i_{N-j+1},{\nu}_{j-2}}&
\rho^r_j(
  {i^\prime_{N-j}},{i^\prime_{N-j+1},{\nu}^\prime_{j-2}};
  i_{N-j},i_{N-j+1},{\nu}_{j-2})
R^j_{(i_{N-j},i_{N-j+1},{\nu}_{j-2}),\nu_j}\nonumber \\
=& &w^r_{\nu_j}
R^j_{
 ({i^\prime_{N-j}},
  {i^\prime_{N-j+1}},
  {\nu}^\prime_{j-2}
 ),\nu_j} \quad ,
\end{eqnarray}
 where we set $\nu_1 \equiv i_{N-1}$.
We can then express the stationary-state vector $|P \rangle$ in
eq.(\ref{eq:def-P}) as
\begin{eqnarray}
& \langle i_0,i_1,\cdots,i_{12},i_{13}
|P \rangle \nonumber\\
 =&
\sum_{\mu_3,\nu_7,\nu_5,\nu_3}
L^3_{(i_0,i_1,i_2),\mu_3}\quad
\psi(\mu_3; i_3,i_4,i_5,i_6; \nu_7) \nonumber\\
&
R^7_{(i_7,i_8,\nu_5),\nu_7}\, 
R^5_{(i_9,i_{10},\nu_3),\nu_5}\, 
R^3_{(i_{11},i_{12},i_{13}),\nu_3}.
\label{eq:bareP}
\end{eqnarray}

The one-point function $\langle i_n \rangle$ can be obtained as
follows. For concreteness, we explain calculation of $\langle i_4
\rangle$ and $\langle i_5 \rangle$.\\

\noindent
[C-1]\\
\noindent
Perform partial summation
\begin{equation}
{\sum_{\{i\}}}^{\prime (4,5)}
\langle i_0,i_1,\cdots,i_{N-2},i_{N-1}|P \rangle
\equiv g(i_4,i_5) \, ,
\end{equation}
 where $\sum_{\{i\}}^{\prime (4,5)}$ means that we do not perform the
summation over $i_4,i_5$.\\

\noindent
[C-2]\\
\noindent
From the definition of  $\langle i_n \rangle$ (eq.(\ref{eq:def-tau})),
\begin{eqnarray}
\langle i_4 \rangle & = &
\left[\sum_{i_5}g(i_4=1,i_5)\right]\Bigm/
\left[\sum_{i_4,i_5}g(i_4,i_5)\right]\\
\langle i_5 \rangle & = &
\left[\sum_{i_4}g(i_4,i_5=1)\right]\Bigm/
\left[\sum_{i_4,i_5}g(i_4,i_5)\right]
\label{eq:i45}
\end{eqnarray}\\
By introducing the following quantities,
\begin{eqnarray}
\sum_{i_0,i_1,i_2} L^3_{(i_0,i_1,i_2),\mu_3}
&\equiv& S^l_3(\mu_3)\\
\sum_{i_{11},i_{12},{i_{13}}}
R^3_{(i_{11},i_{12},i_{13}),\nu_3}
&\equiv & S^r_3(\nu_3)\\
\sum_{i_9,i_{10},\nu_3}
R^5_{(i_9,i_{10},\nu_3),\nu_5} S^r_3(\nu_3)
&\equiv& S^r_5(\nu_5)\\
\sum_{i_7,i_8,\nu_5}
R^7_{(i_7,i_8,\nu_5),\nu_7} S^r_5(\nu_5)
&\equiv& S^r_7(\nu_7)
\end{eqnarray}
we have
\begin{equation}
g(i_4,i_5)
=\sum_{\mu_3,i_3,i_6,\nu_7}
  S^l_3(\mu_3)\, 
  \psi(\mu_3; i_3,i_4,i_5,i_6; \nu_7)\, 
  S^r_7(\nu_7).
\label{eq:g45}
\end{equation}
Eq.(\ref{eq:g45}) means that for calculation of $\langle i_4 \rangle$
 and $\langle i_5 \rangle$ we do not need the full probability
 distribution; we need only $S^l_3(\mu_3), \psi(\mu_3;
 i_3,i_4,i_5,i_6; \nu_7), S^r_7(\nu_7)$.  Accordingly, we need not
 retain full set of arrays
$\{L^j_{({\mu}_{j-2},i_{j-2},i_{j-1}),\mu_j}\}$
 and
 $\{R^j_{(i_{N-j},i_{N-j+1},{\nu}_{j-2}),\nu_j}\}$
 in eq.(\ref{eq:bareP}) at each
 iteration.

\section{Results}
We present the calculated results for two cases: (A) $p=0.75, q=0.25,
\alpha=0.9, \beta=0.8, \gamma=0.01, \delta=0.01$, and (B) $p=0.75,
q=0.25, \alpha=0.5, \beta=0.5, \gamma=0.4, \delta=0.9$. The case (A)
corresponds to the maximal current phase, and (B) to the high density
phase.

In Fig.~\ref{fig-10N-mc}, we compare the DMRG calculation for case (A)
with $ N=10 $ to the exact results (both by the exact diagonalization
and by the program~\cite{HP} using the MPA), varying the number $m$ of
the retained bases. It should be noted that a very small number
($m=2$) of retained bases gives a quite accurate density profile,
implying that the ``weight'' of the density matrix is ``localized''
within very small number of bases. The similar behaviors are seen in
the DMRG calculation for $ N=102 $ in both cases (A) and (B), which
are given in Fig.~\ref{fig-102N-mc} and Fig.~\ref{fig-102N},
respectively. The DMRG calculations with a very small number ($m=2$)
of retained bases are also in excellent agreement with the MPA
results.

It has been known that for a set of parameters, the algebra associated
with the MPA admits finite-dimensional representations.~\cite{HP,MS}
In the $n$-dimensional case, the DMRG with $n$ retained bases should
be exact. The parameters in the present two cases (A) and (B) do not
correspond to such finite-dimensional cases, which means that infinite
$m$ in the DMRG is required for exact calculation. Our results show
that even for such ``infinite-dimensional cases'', DMRG calculation
with a very small number of $m$ serves as an excellent approximation.

\section{Summary}
In this paper, we have applied the density matrix renormalization
group (DMRG) to asymmetric (simple) exclusion process (ASEP) which is
a non-equilibrium many-body problem of hopping particles in one
dimension. We have calculated the density profiles (one-point
function) in the stationary states, which are in excellent agreement
with exact solution. We have found that only a very small number of
retained bases in the DMRG is necessary to attain the close-to-exact
results. We have thus verified that the DMRG is an efficient method to
study the ASEP.

We should comment on possible extensions of the present work.
\begin{enumerate}
\item
Although we have considered only the one-point function in this paper,
we can easily calculate 2-point function $\langle i_m i_n \rangle $,
and, in principle, general $n$-point function in the stationary state. 
This is because, DMRG allows us to obtain the full distribution
function in the stationary state.

\item
The ASEP can be generalized in many different ways. For these
generalized models stationary solutions cannot always be obtained by
the matrix-product-ansatz approach. The DMRG approach used in the
present paper can be easily generalized to these models.  For
instance,
\begin{itemize}

\item
 Models including the processes such as coagulation, decoagulation,
(pair-)creation, and (pair-)annihilation (We can incorporate this
generalization in eq.(\ref{eq:local-TM}).)

\item
 Models of site-dependent parameters of probability $p(n),q(n)$, where
$n$ represents a site in the system

\item
 Models with $N$-species particles ($N\geq2$).

\item
 Models of hopping particles on a ring with a defect. (The DMRG can
deal with 1-dimensional system with periodic boundary condition.)
\end{itemize}

\end{enumerate}

Some of the above are now undertaken, whose results will be given in
future publications.

\section*{Acknowledgments}
The author would like to thank
 Y. Akutsu, K. Okunishi, T. Nishino, M. Kikuchi,
 T. Nagao and S. Yukawa
 for valuable discussions and encouragement.
Part of numerical computation in this work was carried out at the 
 Yukawa Institute Computer Facility.

%
%
%
\newpage
\begin{figure}
    \caption{
        A form of the transfer matrix
        $T=T_4^l {}_\times {}^\times {}_\times T_8^r$
        ($N=14$).
        According to Nishino's diagram~\cite {Nishino},
        $\bigcirc$, $\Box$ and $\Diamond$
        represent
        a single site, a 3-site block and a 7-site block,
        respectively.
}
        \label{fig-14N}
\end{figure}
\begin{figure}
    \caption{The density profile of the model
        in the maximal current phase
        ($N=10, 
        p = 0.75, q = 0.25,
        \alpha = 0.9, \beta = 0.8,
        \gamma = 0.01, \delta = 0.01$)}
        \label{fig-10N-mc}
\end{figure}
\begin{figure}
    \caption{The density profile of the model
        in the maximal current phase
        ($N=102,p = 0.75, q = 0.25,
        \alpha = 0.9, \beta = 0.8,
         \gamma = 0.01, \delta = 0.01$)}
        \label{fig-102N-mc}
\end{figure}
\begin{figure}
    \caption{The density profile of the model
             in the high density phase
        ($N=102,p = 0.75, q = 0.25,
        \alpha = 0.5, \beta = 0.5,
        \gamma = 0.4, \delta = 0.9$)}
        \label{fig-102N}
\end{figure}
%
%
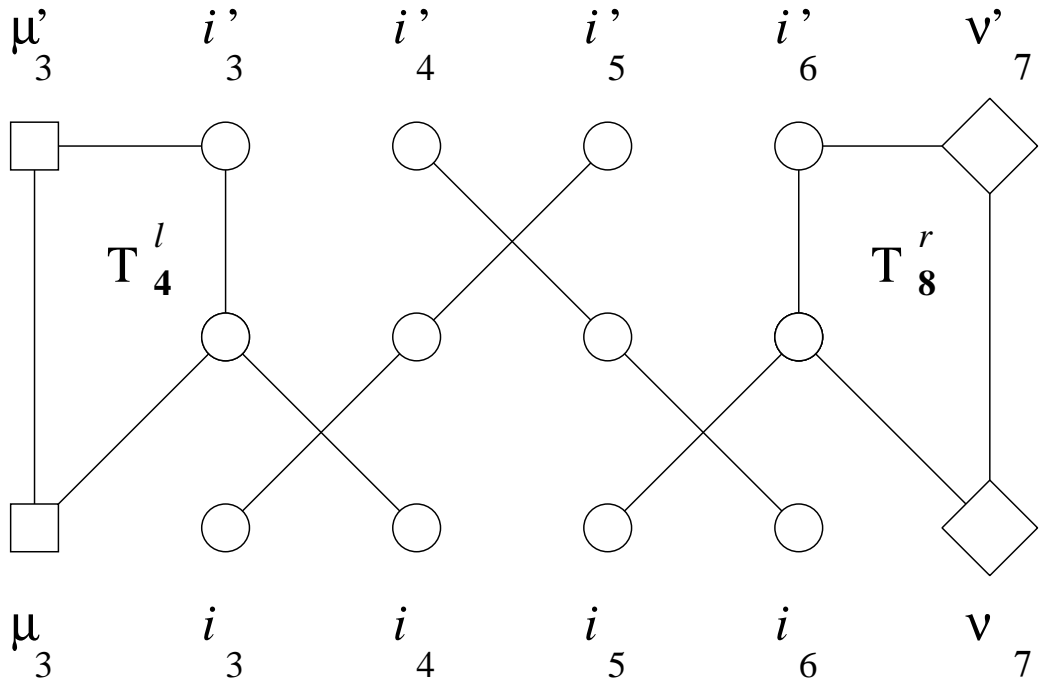
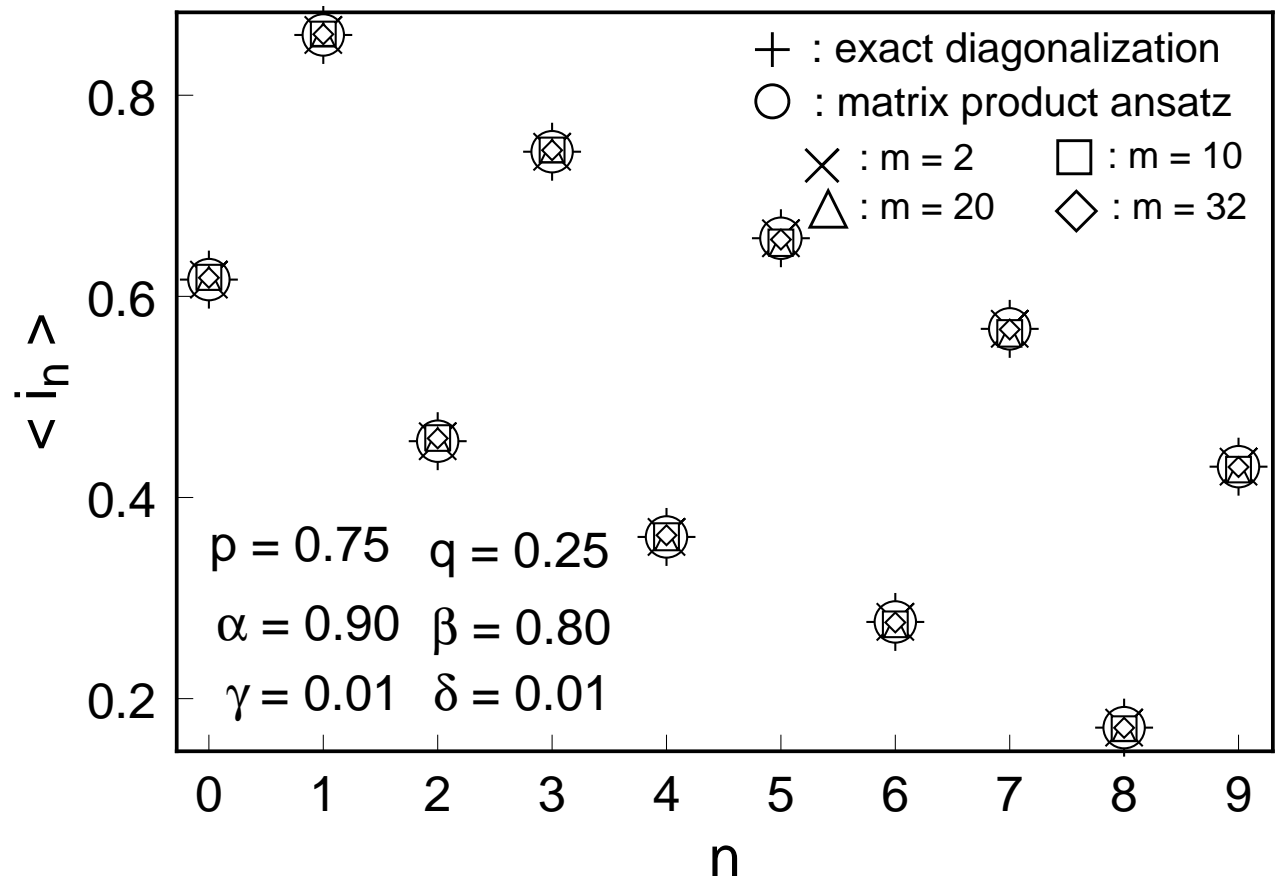
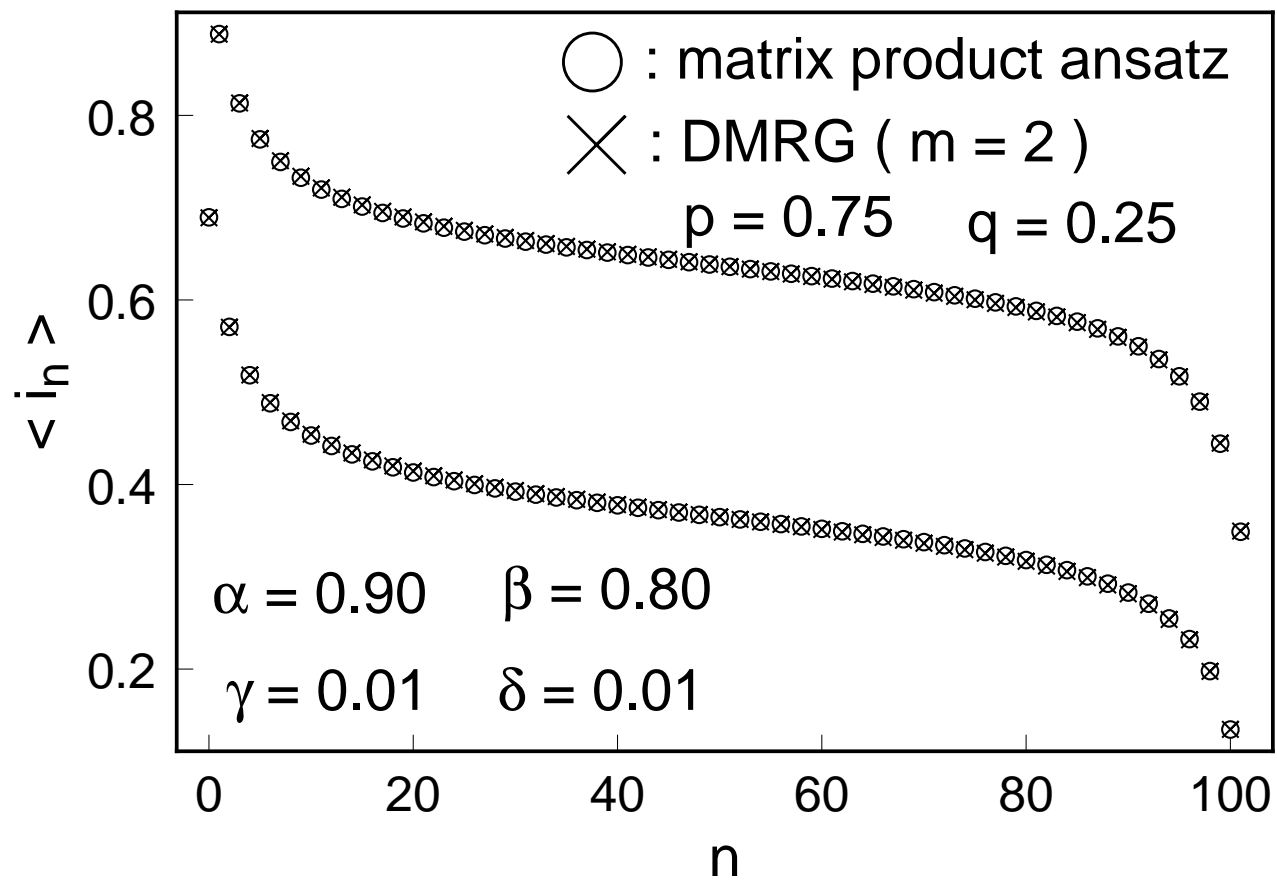
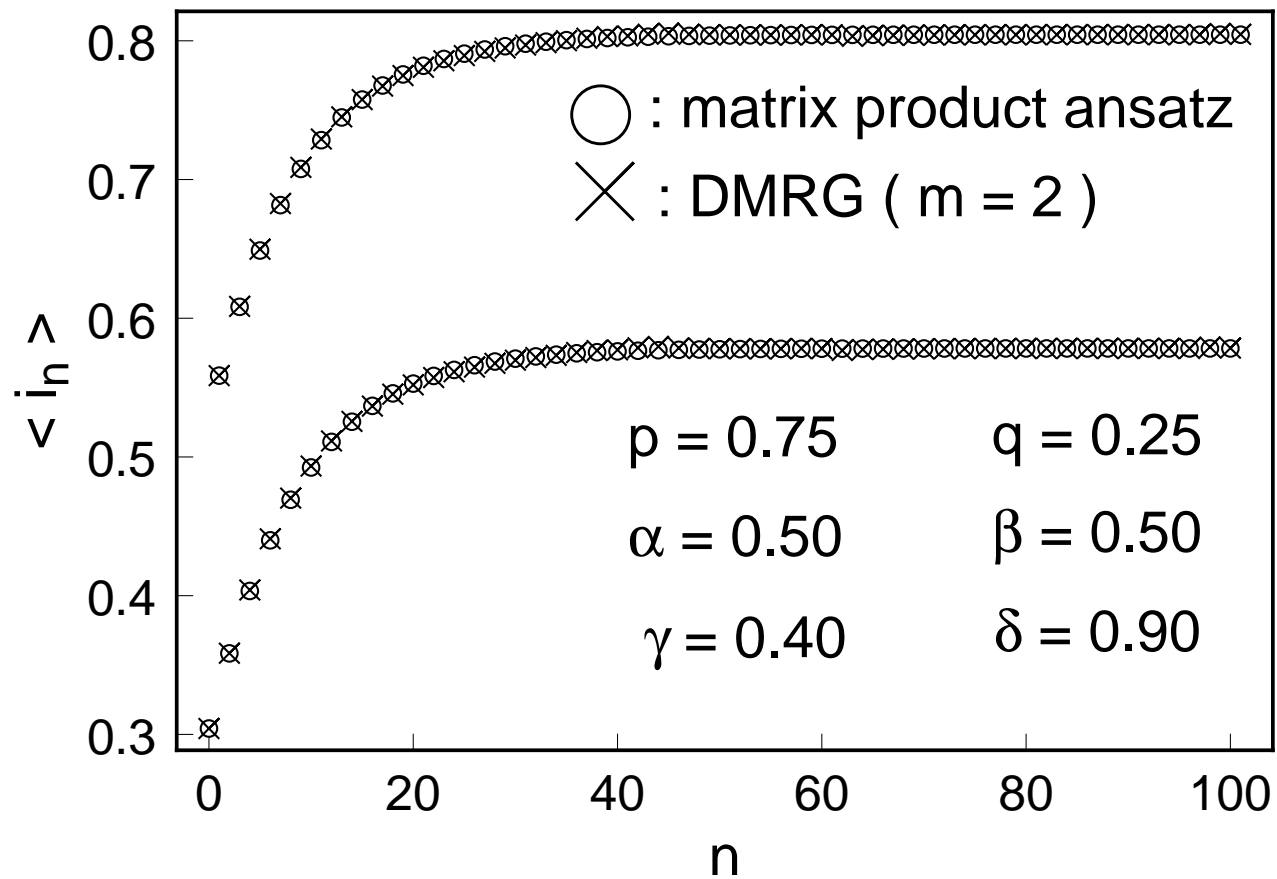
\end{document}